\documentclass[aps,twocolumn,floats,prl,nofootinbib,superscriptaddress]{revtex4}

\usepackage[dvips]{graphicx} %
\usepackage{epsfig,amsmath}
\usepackage{amssymb}

\usepackage{rotate}
\usepackage{color}
\usepackage{aas_macros}

\DeclareFontFamily{OT1}{pzc}{}
\DeclareFontShape{OT1}{pzc}{m}{it}%
            {<-> s * [1.10] pzcmi7t}{}
\DeclareMathAlphabet{\mathscr}{OT1}{pzc}%
                                {m}{it}

\definecolor{RedWine}{rgb}{0.743,0,0}
\definecolor{RoyalBlue}{rgb}{0.25,.41,.88}

\newcommand{\be}{\begin{equation}}
\newcommand{\ee}{\end{equation}}
\newcommand{\bea}{\begin{eqnarray}}
\newcommand{\eea}{\end{eqnarray}}
\def\ba#1\ea{\begin{align}#1\end{align}}

\newcommand{\refeq}[1]{Eq.~(\ref{eq:#1})}          
          
\newcommand{\reffig}[1]{Fig.~\ref{fig:#1}}

\def\bfk{\mathbf{k}}

\newcommand{\DRp}{\ensuremath{\Delta_{{\cal R}0}^2}}
\newcommand{\DR}{\ensuremath{\Delta_{{\cal R}}^2}}

\newcommand{\Mpc}{\ensuremath{\mathrm{Mpc}}}
\newcommand{\kD}{\ensuremath{k_{\mathrm{D}}}}

\newcommand{\me}{\ensuremath{m_{\mathrm{e}}}}
\newcommand{\Omegab}{\ensuremath{\Omega_{\mathrm{b}}}}
\newcommand{\Omegac}{\ensuremath{\Omega_{\mathrm{c}}}}
\newcommand{\id}{\ensuremath{{\mathrm{d}}}}
\newcommand{\expf}[1]{{\rm e}^{#1}}

\newcommand{\keV}{\ensuremath{\mathrm{keV}}}
\newcommand{\MeV}{\ensuremath{\mathrm{MeV}}}
\newcommand{\GeV}{\ensuremath{\mathrm{GeV}}}

\newcommand{\pot}[2]{\ensuremath{#1\times10^{#2}}}

\begin{document}

\title{
Silk damping at a redshift of a billion: \\ 
a new limit on small-scale adiabatic perturbations
}

\author{
Donghui Jeong,
Josef Pradler,
Jens Chluba
and 
Marc Kamionkowski 
}
\affiliation{
Department of Physics and Astronomy, Johns
Hopkins University, 3400 N.\ Charles St., Baltimore, MD 21218
}

\begin{abstract}
 We study the dissipation of small-scale adiabatic perturbations at early times
 when the Universe is hotter than $T \simeq 0.5\,\keV$. When the wavelength falls
 below the damping scale $\kD^{-1}$, the acoustic modes diffuse and thermalize, 
 causing entropy production. Before neutrino decoupling, $\kD$ is primarily 
 set by the neutrino shear viscosity, and we study the effect of acoustic 
 damping on the relic neutrino number, primordial
 nucleosynthesis, dark-matter freeze-out, and baryogenesis. This
 sets a new limit on the amplitude of  
 primordial fluctuations of $\Delta_{\mathcal R}^2 < 0.007$ at 
 $10^4 \,\Mpc^{-1}\lesssim k\lesssim 10^5\,\Mpc^{-1}$ and a model dependent 
 limit of $\Delta_{\mathcal R}^2 \lesssim 0.3$ at 
 $k \lesssim 10^{20-25}{\rm Mpc}^{-1}$.
\end{abstract}
\date{\today}

\maketitle

\paragraph{Introduction. }
A wealth of astronomical observations, especially measurements of the 
cosmic microwave background (CMB) temperature and polarization anisotropies,
have elevated the hot Big Bang to a detailed and precise model
for the early Universe. There is also strong evidence that the
Universe underwent a period of inflationary 
expansion which sets the initial conditions for the growth of
large-scale structure from primordial curvature perturbations.
Given the initial conditions, a simple model of a flat Universe that is 
filled with baryons, cold dark matter, neutrinos and a cosmological constant 
($\Lambda$CDM), already describes the data extraordinarily well
\citep{WMAP_params, Planck2013params}.

The success of $\Lambda$CDM on large scales is reflected in the
precise determination of cosmological parameters, such as the
contributions of baryons, $ \Omegab h^2 = 0.02205\pm0.00028 $, and
cold dark matter (DM), $\Omegac h^2 = 0.1199 \pm 0.0027$, to the universal
energy density, or the effective number $N_{\rm eff} =3.30\pm
0.27$ of massless neutrinos $\nu$ \citep{Planck2013params}. The latter
informs us about the particle content at matter-radiation equality and
is currently consistent with that of the Standard Model  (SM)
of particle physics. It is furthermore common belief that a Universe, 
endowed with such minimal field content, had an ``uneventful'' thermal 
history between the epochs of big-bang nucleosynthesis (BBN)---or possibly 
even between dark-matter freeze out (FO)---and hydrogen
recombination, so that the number-to-entropy ratios remain constant, $(N_{b,c,\nu} - N_{\bar
  b,\bar c ,\bar \nu} )/S|_{\rm CMB} = (N_{b,c,\nu} - N_{\bar b, \bar
  c ,\bar \nu })/S|_{\rm BBN/FO}$.  Importantly, this allows one to
perform cosmological concordance tests from BBN light-element yields,
to judge the viability of DM models from their expected FO abundance,
to infer parameters for successful baryogenesis, or to contemplate or
discard extensions of the SM $\nu$ sector.

In this \textit{Letter}, we emphasize that the above rationale
carries the implicit assumption $\int \Delta_{\cal R}^2(k) \id
\ln k \ll 1$, where $\Delta_{\cal R}^2(k)$ is the variance of
the primordial curvature amplitude on wavelengths $k$.
This is because a fraction $\delta \rho/\rho \propto \zeta$ of the total energy 
density is stored in the primordial curvature perturbations $\zeta$.
Once a mode with wavenumber $k$ enters the
horizon, it becomes dynamical (an ``acoustic wave'') and dissipates its
energy by particle diffusion, commonly referred to as Silk
damping~\cite{Silk1968} when it regards the photon-baryon fluid in the
post-BBN era. This process leads to entropy production, or more concisely 
changes in the particle number, and consequently affects the early thermal 
history.

While the amplitude $\Delta^2_{\cal R}(k) \equiv
\left<|\zeta|^2\right> \simeq \mathcal{O}(10^{-9})$ of
the primordial power spectrum at scales
$10^{-3}\,\Mpc^{-1}\lesssim k \lesssim 3\,\Mpc^{-1}$
is tightly constrained by the
CMB~\cite{Hinshaw2013,Planck2013params}, galaxy clustering
\cite{Sanchez2013}, and the Lyman-$\alpha$ forest
\cite{BPVV11}, $\Delta_{\cal R}^2$ remains essentially unconstrained on
smaller scales (larger $k$). Upper limits at $k\gtrsim 3\,\Mpc^{-1}$ available 
in the literature are derived from limits on CMB spectral distortions
\citep{Fixsen1996, Hu1994, Chluba2012, Chluba2012inflaton,
Chluba2013iso}, the absence of evidence for primordial black holes
\cite{JGM09}, and from indirect constraints of DM annihilation inside
ultra-compact mini-halos \cite{BSA11}.
Here, we add an independent constraint for $k \gtrsim 10^4\,\Mpc^{-1}$ that 
can be viewed as more robust in that it derives directly from an altered 
thermal history of the early Universe and is independent of any new physics 
beyond the SM.

Our work expands on earlier investigations that primarily discuss the
damping of perturbations at low redshift $z\lesssim 2\times10^6$ 
(the spectral-distortion era), where energy injection from
dissipation is not fully thermalized but rather leads to a
readjustment of the photon spectrum
\cite{Zeldovich1969,Sunyaev1970SPEC, Chluba2012}.
At high redshift $z\gtrsim2\times 10^6$ (the blackbody era),
photon-number--changing interactions quickly restore a
blackbody spectrum, so that any direct observable from the CMB is wiped
out \citep{Sunyaev1970mu, Danese1982, Burigana1991, Hu1993, Chluba2011therm}.
Therefore the blackbody era has received little attention in the
past.  As we show below, though, early energy release modifies the thermal
history of the Universe at $T\gtrsim{\rm keV}$ and thus 
the standard calculations of neutrino number, BBN, baryon-to-photon
ratio $\eta_b$, and dark-matter relic density.

\paragraph{Dissipation of acoustic modes.}
Let us denote the total energy density of relativistic particles
in equilibrium with photons as $\rho=\sum \rho_i$ and the
energy density of individual species $i$ by $\rho_i$. Similarly, we
write $N=\sum N_i$ for the average number density of particles.
For adiabatic initial conditions, the photon density perturbations
outside the horizon $\delta_\gamma^i(\bfk)=\delta
\rho_\gamma/\rho_\gamma$ are related to the primordial curvature
perturbation $\zeta(\bfk)$ by $\delta_\gamma^i(\bfk) =
-(4/3)\,C\,\zeta(\bfk)$ where $C=1$ and $C= (1+4/15 R_{\nu})^{-1}$
before and after neutrino decoupling, respectively~\cite{Ma1995};
$R_\nu \equiv \rho_\nu/(\rho_\nu + \rho_\gamma)$; and we assume
neutrino decoupling as instantaneous at temperature
$T_{\nu,\rm dec}=1.5\,\MeV$.
After entering the horizon, radiation-density perturbations evolve
as $\delta_\gamma(t, \bfk) \approx 3 \delta_\gamma^i(\bfk) \cos\left[k
  r_s(t)\right]\exp[-k^2/\kD^2(t)]$ where $r_s$ is the sound horizon at
time $t$~\cite{Hu1995CMBanalytic, WeinbergBook}, and $\kD(t)$ is the diffusion
scale below which ($k = | \bfk | >\kD$) modes are being dissipated. 

The presence of primordial perturbations implies an universal average
photon energy and number density of $\rho_\gamma\simeq a_{\cal B}
\bar{T}^4(1+6\left<\Theta^2\right>)$ and $N_\gamma\simeq b_{\cal B}
\bar{T}^3(1+3\left<\Theta^2\right>)$. Here, $\bar{T}=\left<T\right>$
is the average temperature of the Universe, and $\Theta(t, {\mathbf x},
{\hat n})=\Delta T/\bar{T}$ denotes the local temperature perturbation
at some fixed time $t$ in different directions $\hat n$. The
angle brackets $\left<...\right>$ denote averages over space at
some fixed time.
In comparison, a \textit{blackbody} at temperature $ T$ has
$\rho_{\gamma}=a_{\cal B} T^4$ and $N_{\gamma}=b_{\cal B} T^3$. From
this, one finds that the presence of perturbations is associated
with a momentary lack of photons, $\Delta N_\gamma/N_\gamma\approx
(3/2) \left<\Theta^2\right>$ (which will be replenished by the thermalization
process) and a corresponding excess energy density, $Q_\gamma\simeq
2 \rho_\gamma \left<\Theta^2\right>$.  A similar picture holds for
any other species contributing to $\rho$ defined above.

In the CMB rest frame and at sub-horizon scales, we have
$\left<\Theta^2\right>\simeq
\left<\Theta_0^2\right>+3\left<\Theta_1^2\right>\simeq
\left<|\Theta_0|^2\right>$, where we used that in the tight-coupling
regime the amplitude of the photon dipole is $|\Theta_1|\simeq
|\Theta_0|/\sqrt{3}$ and $\pi/2$ out of phase with the monopole
\cite{Hu1995CMBanalytic};
in Fourier space, $|\Theta_0|_k \approx (3/4)\delta^i_\gamma
\exp(-k^2/\kD^2)=C \zeta \exp(-k^2/\kD^2)$. Assuming adiabatic
perturbations, $\delta\rho_i/\rho_i \simeq \delta
\rho_\gamma/\rho_\gamma$ for all $i$, the average fractional energy
release (from the acoustic waves to the average plasma) between time
$t_1$ and $t_2$ is then given by~\cite{Chluba2012, Chluba2013iso}
\begin{align}
  \label{eq:Drho_ov_rho}
  \frac{\Delta Q}{\rho} 
  &\approx 2\left[\left<\Theta^2\right>_{t_1}\!-\!\left<\Theta^2\right>_{t_2}\right]
  \approx 2C^2\!\!
  \int_{\kD(t_2)}^{\kD(t_1)}\!\frac{\id k}{k}  \Delta_\mathcal{R}^2(k).
\end{align}
Here, $\Delta_{\mathcal{R}}^2(k)$ is related to the primordial
curvature power spectrum by $\Delta_{\mathcal{R}}^2(k)\equiv k^3
P_{\zeta}(k)/(2\pi^2)$. 

For $z\gtrsim2\times 10^6$, the energy release above yields 
the entropy production, or the change in comoving number density of 
relativistic particles, as 
$\id \ln a^3 N/\id t \approx -(3/2) \partial_t \left<\Theta^2\right>$,
from which we calculate the photon number density as
\ba
\label{eq:Tz}\!
N_\gamma(z) \approx N_\gamma^\ast(z) \exp\!
\left[-\frac{3 C^2}{2}  \!\! \int_{0}^z
 \!\! \DR(\kD) \frac{d\ln\kD}{d\ln z} d\ln z\right].
\ea
Note that similar relations hold for all relativistic
particles thermally coupled to photons.
Here, $N_\gamma^\ast(z)$ is the average photon number without
thermalization but taking into the account the smoothing of perturbations by 
particle diffusion. That is, $N_\gamma^\ast(z)$ is the photon number
density at redshift
$z$ extrapolated from the CMB temperature today $T_0=2.726~{\rm K}$ and 
the standard thermal history including the entropy transfer from $e^{\pm}$
annihilation, etc. Eq.~(\ref{eq:Tz}) defines an effective
photon temperature, $T\equiv (N_\gamma/b_{\cal B})^{1/3}$, specific
to the average number of photons in the Universe, and 
$T^\ast\equiv (N_\gamma^\ast/b_{\cal B})^{1/3}$, the temperature that appears 
in the usual themal history calculation.
At this point, it is worth stressing that the total radiation energy density 
in the Universe stays practically unchanged throughout the diffusion and 
thermalization process, because it simply redistributes the energy 
stored inside of the perturbations to the median and the thermalization 
only changes particle numbers, but not the energy density.
As a result, the expansion history at early times is the same as the usual
calculation and controlled by $T^\ast(z)$.

\paragraph{Diffusion scale.}
\begin{figure}[tbp]
\hspace{-4.4mm}\includegraphics[width=1.05\columnwidth]{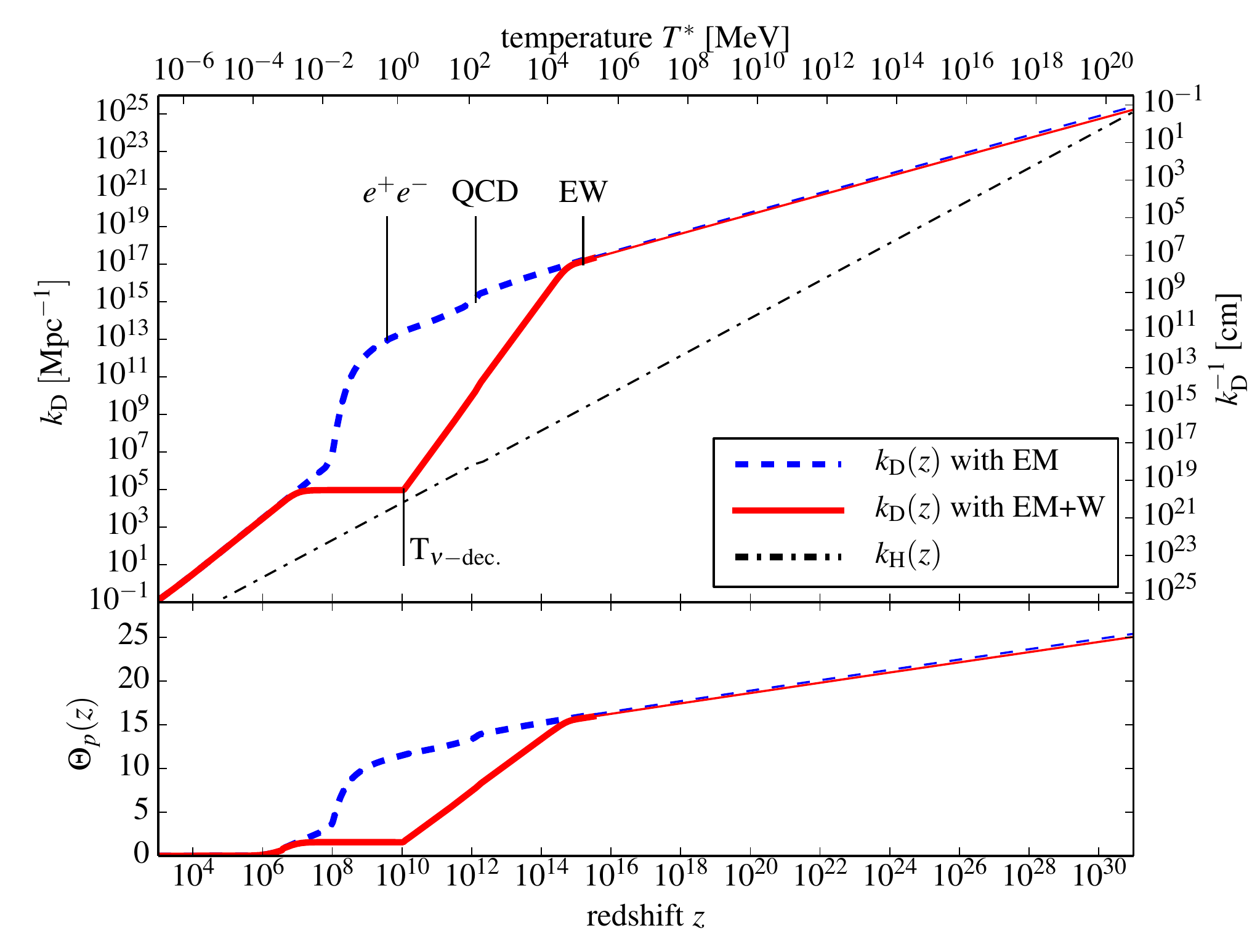}
\caption{Redshift and temperature dependence of the diffusion
    scale $\kD$ (top) and temperature parameter $\Theta_p$
    (bottom) defined in \refeq{Thetap}. The blue dashed line captures photon 
    diffusion (without neutrino shear viscosity); the red
    solid line includes neutrino diffusion and represents the full
    result.  Neutrino shear viscosity dominates dissipation before
    neutrino decoupling, with a diffusion scale that is close to 
    the comoving horizon $k_{\rm H} = aH$(black dot-dashed line).
    Lines at $T>200{\rm GeV}$ are thin, as that
    part of the graph may be modified if there are particles
    or interactions beyond the SM.
} \label{fig:kD}
\end{figure}

We calculate $\kD$ from the damping rate $\Gamma(k,t)$, $\kD^{-2} =
k^{-2}\int_0^t \id t'\Gamma(k,t')$. At early times, heat conduction and bulk
viscosity of the plasma are negligible~\cite{Weinberg1971}, and
$\Gamma$ is dominated by shear viscosity $\eta$, 
i.e., $\Gamma(k,t) \approx \frac23 \frac{k^2}{a^2(\rho+p)} \eta(t)$.
Here, $p\simeq \rho/3$ is the pressure of the primordial fluid and
$\eta$ is roughly given by~\cite{Weinberg1971,Kaiser1983}
\begin{align}
\label{eq:eta}
  \eta = \frac{16}{45}\rho_{\gamma}t_{\gamma} + \frac{4}{15}\rho_{\nu
    }t_{\nu}\Theta(T-T_{\nu, \rm dec}).
\end{align}
$t_{\gamma} = (n_{e^{\pm}} \sigma_{\rm KN})^{-1}$ denotes the mean
free scattering time of a photon with $n_{e^{\pm}} $ being the electron-positron
number density and $\sigma_{\rm KN}(x)$ the Klein-Nishina cross
section for which we use the expression
in~\cite{Rybicki1979} with $x = 2.7 \frac{T}{\me} \left( 3
  \frac{T}{\me} + \frac{K_1(\me/T)}{K_2(\me/T)} \right)$ as a thermal
averaged quantity; and $K_{\nu}$ is a Bessel function of the second kind.
Before neutrinos stream freely, their mean free time $t_{\nu}$ is
determined by weak interactions with $\sigma_{\nu} \approx (G_F T)^2 =
5.3\times10^{-44} \,T_{\rm MeV}^2~{\rm cm}^2$. We also estimate $\kD$
at temperatures above the electroweak phase transition $T_W =
O(100\,\GeV)$ from electroweak interactions of the SM content. For
$T\gg T_W = 100\,\GeV$ the temperature dependence of the scattering
cross section becomes one of a gauge interaction $\sigma_{\nu} \propto
T^{-2} $ and, for simplicity, we assume an instant (second order)
phase transition and take $Z$ and $W$ bosons as massless for
$T>T_W$. A more detailed numerical study will neither change the
qualitative picture nor the quantitative analysis by much.
At low redshift, $z \lesssim 10^8$, and with the inclusion of heat
conduction $\chi$ in $\Gamma$, the damping rate reduces to the
expression familiar in the CMB literature~\cite{WeinbergBook}.

As a general rule, the particle that is most weakly interacting, yet still
kinetically coupled and as abundant as radiation, controls $\kD$.
This is because it will have the largest product $t_{i} \rho_i$ in a
generalization of Eq.~(\ref{eq:eta}) for $\eta$. For the purpose of
this work we assume a SM field content and a massive DM particle with
an electroweak-strength interaction. It is then the massless SM
degrees of freedom which suffice to be taken into account for
calculating~$\kD$.

Importantly, from Eq.~(\ref{eq:Drho_ov_rho}) we see that $\kD(z)$
informs us about the scales $k$ that dissipate at a given
redshift~$z$. 
The redshift evolution of the diffusion scale is shown as a red, solid
line in \reffig{kD}.  The diffusion scale at $T>T_W$ (where
anisotropic shear from $\gamma$, $W^\pm$, $Z$ bosons are all
important), $\kD\simeq 4.5\times 10^{14} (T/{\rm MeV})^{0.51}~{\rm
  Mpc}^{-1}$.  After electroweak symmetry breaking and before neutrino
decoupling, $T_{\nu,\rm dec}<T<T_W$, $\kD$ is dominated by neutrino
shear viscosity, $\kD\simeq 5\times10^4(T/{\rm MeV})^{2.7}~{\rm
  Mpc}^{-1}$.  For $T<T_{\nu, \rm dec}$ , $\kD\simeq 10^5~{\rm Mpc}^{-1} $
remains constant until $T\simeq 2\,\keV$. 
 This is because neutrinos previously erased near-horizon sized modes
  (dot-dashed line) so that the uptake of photon diffusion is delayed
  until a later epoch at $T\simeq 2\,\keV$ since
  $t_{\gamma} \ll t_{\nu} $.  This makes additional photon production
  almost negligible during the epoch of BBN.

\paragraph{Revised thermal history.}
In the spectral-distortion era, the limits on
$\Delta_{\mathcal{R}}^2(k) $ from $\mu$ distortions of the CMB are
already quite stringent~\cite[e.g.,][]{Chluba2012inflaton} and photon
heating is not relevant.  Therefore, we focus on dissipation in the
thermalization era ($z>2\times 10^6$, $k\gtrsim 10^4~{\rm Mpc}^{-1}$).
For simplicity, let us assume that the amplitude of primordial
curvature fluctuations is scale-invariant on small scales with
amplitude $\DRp$.  Thus, the average photon temperature becomes 
\begin{align}
  T(z) = T^\ast(z)\, \expf{-\DRp\Theta_p(z)}.
  \label{eq:Thetap}
\end{align}
The definition of $\Theta_p(z) $ can be deduced from
Eqs.~\eqref{eq:Drho_ov_rho} and \eqref{eq:Tz} [see the bottom panel of
\reffig{kD}].  The plateau value for $2\,\keV < T< T_{\nu, \rm dec}$
is $\Theta_p \simeq 1.6$.  We find that $\Theta_p(T^\ast) \simeq
1.3\ln(T^\ast/{\rm MeV}) + 1.3 \simeq 1.3\ln z - 30$ and
$\Theta_p(T^\ast) \simeq 0.25\ln(T^\ast/{\rm MeV}) + 13 \simeq 0.25\ln z
+ 7.2$ are good approximations for, respectively, $T_{\nu,\rm
  dec}<T<T_W$ and $T>T_W$. Equipped with the modification of the
$T$-$z$ relation, we shall now discuss its consequence for
cosmological observables.

\paragraph{Neutrino number density and $N_{\rm eff}$.}
After neutrino decoupling, the comoving neutrino number remains
constant. However, photon production from dissipation of acoustic
waves continues and $N_\nu/N_\gamma$ changes with time. At the same
time, it is important to note that $N_{\rm eff}$, which measures
the {\it energy density} of relativistic particles, remains fixed at its
standard value.  This is because both the neutrino and photon fluid
initially shared the same perturbations and energy conservation
implies that their relative energy densities are not affected by the
presence and dissipation of small-scale modes.
Albeit strictly beyond the scope of present experimental capabilities,
we note in passing that any direct observation of $N_{\nu}$ in
mismatch with a value inferred from $N_{\rm eff}$ can in principle
allow us to probe the the small-scale power spectrum at $k\lesssim
10^5\,\Mpc^{-1}$. Further information may then be extracted from {\it
  neutrino spectral distortions}, which are caused by mixing of 
Fermi-distributions of slightly different temperature in the neutrino 
free-streaming phase.

\paragraph{Light element yields.}
As alluded to before, no entropy is produced after neutrino decoupling
until $T\simeq 2\,\keV$. This frames the period of nucleosynthesis and the
modifications to BBN come from an elevated baryon asymmetry as initial
condition (because of post-BBN dissipation of small scale power) and a
modification of the average energy per particle, $\rho/N$.

\begin{figure}[tb]
\includegraphics[width=\columnwidth]{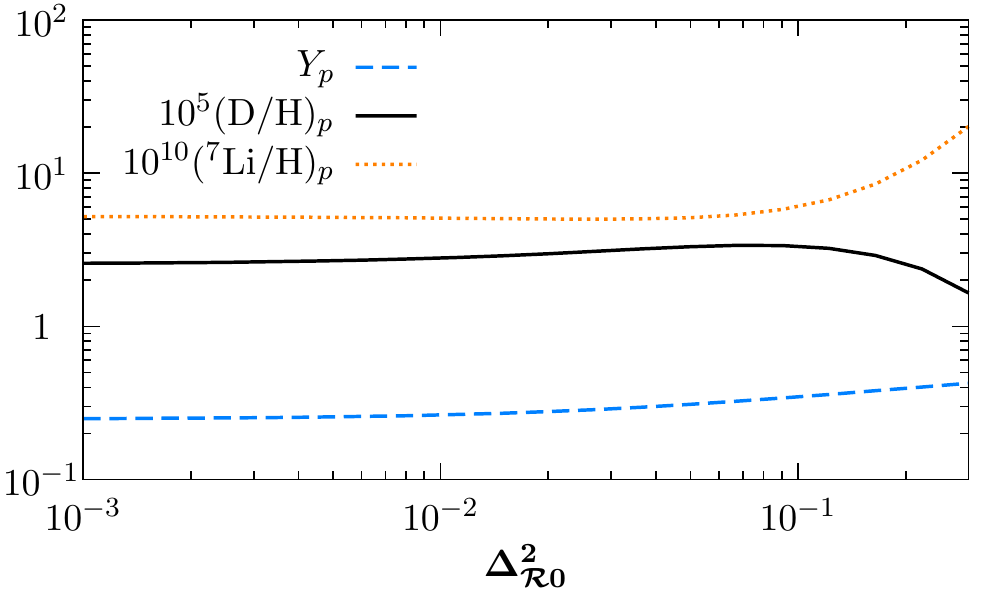}
\caption{Light-element yields from BBN as a function of
$\DRp$. The dominant constraint is derived from an
overproduction of helium.
\label{fig:bbn2}
}
\end{figure}

Impressive progress has been made in the determination of the
primordial deuterium abundance from high-$z$ QSO absorption systems,
with the most recent mean reported as $({\rm D/H})_p = (2.53 \pm 0.04)
\times 10^{-5} $~\cite{Cooke2014}. A precision measurement of the
true primordial He abundance must await future CMB probes; inference
of the primordial mass fraction $Y_p$ from extragalactic H-II regions
are plagued by systematic
uncertainties~\cite{Izotov2010,Aver2013} and a conservative
range may be taken as~$0.24\leq Y_p \leq 0.26$.
The constraint applies to those modes that dissipate their energy
\textit{after} BBN but before the spectral-distortion era, $k\simeq10^4 -
10^5~{\rm Mpc}^{-1}$. We find, 
\begin{align}
\label{eq:bbn}
   Y_p :\, \DRp< 0.007, \quad ({\rm D/H})_p:\, \DRp< 0.2 , 
\end{align}
from the overproduction of He; for D/H we adopted a nominal
$2\sigma$ lower limit from the quoted mean. 
Since higher values of primordial D/H are in principle conceivable
(e.g., by systematic D absorption on dust grains), we refrain
from deriving a limit on the overproduction of D/H. However, no known
astrophysical sources of D exist, and underproducing D yields a robust
constraint, Eq.~(\ref{eq:bbn}). We note that Li/H increases
with larger $\DRp$, worsening the cosmological lithium problem 
(see Fig.~\ref{fig:bbn2}).

\paragraph{Baryon asymmetry and DM relic abundance.}
Entropy production causes a dilution of net particle number once the
latter is frozen out. We have already seen the reduction of
$\eta_b$ in the post-BBN era. Here, we will consider the entropy
conversion from $\kD$ evolution for temperatures well above
$1\,\MeV$. Our results are derived under the premise that $\kD$ is
governed by SM fields only. 

The photon production from dissipation of acoustic modes dilutes the
baryon-to-photon ratio,
\be \ln\left(\frac{N_B}{N_\gamma}\right) \approx \ln\left(
  \left.  \frac{N_B}{N_\gamma} \right|_\ast \right) + 3 \DRp
\Theta_p(T^\ast).  
\ee 
Therefore, in the presence of small-scale power, the initial baryon
asymmetry from some baryon-number-generating process has to be larger
than in the standard case.
Above the QCD phase transition, baryon number is carried by quarks so
that $N_B\simeq N_\gamma$, implying a principal limit
$(N_B-N_{\bar{B}})/N_\gamma \lesssim \mathcal{O}(1)$. The latter condition implies
\be \DRp \lesssim 21 \left[ 39 + 28 \ln \left(\frac{T}{10^{19}\rm
      GeV}\right) \right]^{-1}.  \ee
For baryogenesis scenarios operative at $T\sim {\rm TeV}$ to
$10^{19}{\rm GeV}$, this gives a rather weak bound $\DRp \lesssim
0.3$. However, it must be said that the constraint applies to
remarkably small scales, $\kD\simeq 10^{20-25}~{\rm Mpc}^{-1}$.

\begin{figure}[tbp]
\includegraphics[width=0.49\textwidth]{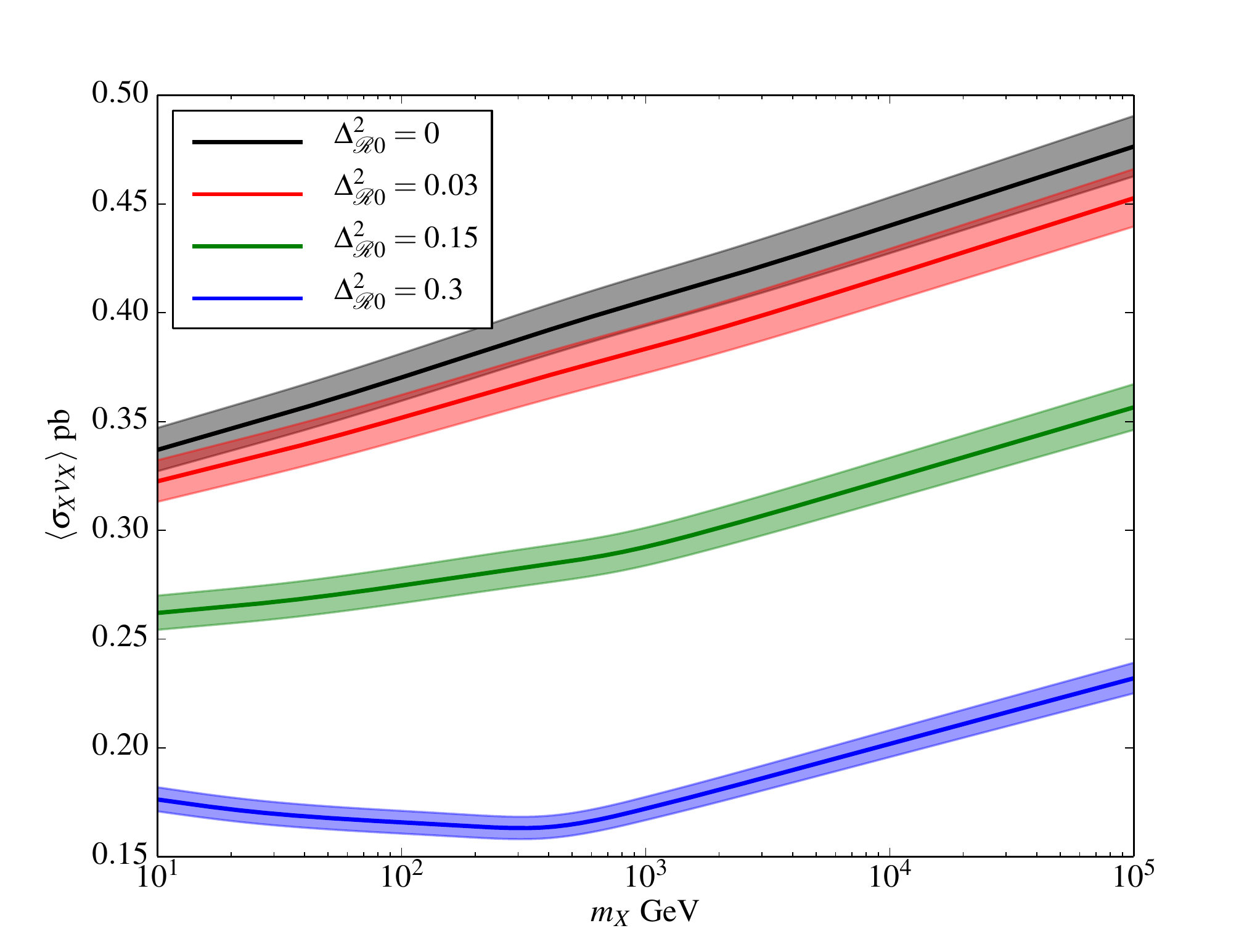}
\caption{Estimate of the DM annihilation cross section $\left<\sigma v\right>$  in
  the presence of small-scale density perturbations as a
  function of DM mass $m_\chi$.
 The bands  indicate the region $\Omega_{\rm c} = 0.2594 \pm 0.0074$
 (95\% C.L. from \cite{Planck2013params}).
\label{fig:Thetap}
}
\end{figure}

\paragraph{DM relic abundance.}
  The calculation of the dark-matter relic abundance is also
  affected by a revised temperature-redshift relation. If DM is a
  weakly-interacting massive thermal relic, its abundance freezes out
  when the annihilation rate equals the expansion rate: $H \simeq n_{\rm
    DM}\left<\sigma v\right>$. For given $H$, the DM equilibrium
  number density is reduced relative to the standard case, leading to
  an earlier FO. Conversely, the expansion factor from FO to
  the present increases by a factor of $\expf{3\DRp\Theta_p}$, and that
  reduces the relic DM number density. The values $\left<\sigma
    v\right>$, required for matching onto the CMB observation of
  $\Omega_{\rm c} $, are shown in a sample calculation in
  Fig.~\ref{fig:Thetap}.

\paragraph{Conclusion.}
We study the dissipation of primordial acoustic waves from adiabatic
perturbations, and its impact on the thermal history of the early
Universe at redshift $z\gtrsim \pot{2}{6}$.  
Because of dissipation, the redshift-temperature relation is modified
and entropy 
production leads to a revision of $N_\nu/N_\gamma$, (D/H)$_p$,
$Y_p$. From those observables we establish a constraint $\DRp < 0.007$
at comoving scales $10^4\,\Mpc^{-1} \lesssim k \lesssim
10^5\,\Mpc^{-1}$.
Such small scales were previously believed to be inaccessible by
\textit{direct} early Universe observables. 

One can take this work into various directions that remain to be
explored. For example, we restricted ourselves to a SM particle
content. New radiation degrees of freedom that are populated for
$T>1\,\MeV$ and that have interaction strengths such that their mean
free path exceeds the one of neutrinos are likely to dominate the
plasma's viscosity. This can lead to more drastic modifications of the
thermal history prior to BBN, with consequences for baryogenesis and
the DM problem. Even within SM with massive neutrinos, the diffusion
scale at high $T$ will be model dependent. For example, say, neutrinos
are Dirac particles, and their right-handed counterparts are fully
excited for temperatures well above their mass. They may then dominate
the diffusion process when the only link to the thermal bath comes
from minute Yukawa interactions.
 
We have also only considered Gaussian primordial fluctuations and
wavemodes that are statistically independent. However, if long- and
short-wavelength fluctuations are correlated (e.g., through
local-model non-Gaussianity), the dissipation on small scales
will give rise on large scales to an isocurvature fluctuation
correlated to the adiabatic perturbation.  We leave the study of
the effects of these modes to future work. 

\small
\smallskip
\paragraph{Acknowledgments.}  This work was supported by NSF
Grant No.\ 0244990 and by the John Templeton Foundation.

\bibliography{Lit}

\end{document}